\documentclass[11pt,titlepage, a4paper]{article}
\usepackage{amsthm}
\usepackage{mathptmx}       
\usepackage{helvet}         
\usepackage{courier}        
\usepackage{graphicx}        
\usepackage[bottom]{footmisc}
\usepackage{endnotes} 
\makeindex 
\theoremstyle{definition}

\newtheorem*{ann}{Fundamental assumption for cognitive and brain sciences}
\newtheorem*{decoan}{Reverse engineering the cortex}

\begin{document}

\title{On Reverse Engineering in the Cognitive and Brain Sciences
} 

\author{Andreas Schierwagen\\
\\Institute for Computer Science, Intelligent Systems Department\\ University of Leipzig\\ Leipzig, Germany\\
\\schierwa@informatik.uni-leipzig.de\\
}

\maketitle      

\begin{abstract}
 Various research initiatives try to utilize the operational principles of organisms and brains to develop alternative, biologically inspired computing paradigms and artificial cognitive systems. This paper reviews  key features of  the standard method applied to complexity in the cognitive and brain sciences, i.e. decompositional analysis or reverse engineering. The indisputable complexity of brain and mind raise the issue of whether they can be understood by applying the standard method. Actually, recent findings in the experimental and theoretical fields, question central assumptions and hypotheses made for reverse engineering. \\
Using the modeling relation as analyzed by Robert Rosen, the scientific analysis method itself is made a subject of discussion. It is concluded that the fundamental assumption of cognitive science, i.e. complex cognitive systems can be analyzed, understood and duplicated by reverse engineering, must be abandoned. Implications for investigations of organisms and behavior as well as for engineering  artificial cognitive systems are discussed.
\end{abstract}

\section{Introduction}
\label{sec:1}

For some time past, computer science and engineering devote close attention to the functioning of the brain. It has been argued that recent advances in cognitive science and neuroscience have enabled a rich scientific understanding of how cognition works in the human brain. Thus, research programs have been initiated by leading research organizations to build new computing systems based on information processing principles derived from the working of the brain, and to develop new cognitive architectures and computational models of human cognition (see, e.g. \cite{Sch2007,Sch2009}, and references therein). 

Two points are emphasized in those research programs:  First, there is impressive abundance of available experimental brain data, and second, we have the computing power to meet the enormous requirements to simulate a complex system like the brain. Given the improved scientific understanding of the operational principles of the brain as a complexly organized system, it should then be possible to build an operational, quantitative model of the brain. Tuning the model could be achieved then using the deluge of empirical data, due to the ever-improving experimental techniques of neuroscience. 

Trying to put this idea into practice, however, has generally produced disenchantment after high initial hopes and hype. If we rhetorically pose the question ``What is going wrong?'' (as previously posed in the field of robotics \cite{Brooks2001}), possible answers are: (1) The parameters of our models are wrong; (2) We are below some complexity threshold; (3) We lack computing power; (4) We are missing something fundamental and unimagined. In most cases, only answers (1)-(3) are considered by computer and AI scientists, and allied neuroscientists, and conclusions are drawn in similar vein. If  answer (1) is considered true, still better experimental methodologies are demanded to gather the right data, preferably at the molecular genetic level (e.g. \cite{Novere2007}). Answers (2) and (3) often induce claims for concerted, intensified efforts relating phenomena and data at many levels of brain organization (e.g. \cite{Grillner2005}).

Together, any of answers (1)-(3) would mean that there is nothing {\it in principle} that we do not understand about brain organization. All the concepts and components are present, and need only to be put into the model. This view is widely taken; it represents the belief in the efficiency of the scientific method, and it leads one to assume that our understanding of the brain will major advance as soon as the `obstacles' are cleared away. 

As I will show in this paper, there is, however, substantial evidence in favour of answer (4). I  will argue that, by following the standard scientific method, we are in fact ignoring something fundamental, namely that biological and engineered systems are basically different in nature.

The paper is organized as follows. Section \ref{sec:2} presents conceptual and methodological basics of the cognitive and brain sciences. The concepts of decompositional analysis and localization underlying the reverse engineering method are reviewed. I discuss the idea of modularization and its relation to the superposition principle of system theory. Then, Section \ref{sec:5}  shortly touches on Blue Brain and  SyNAPSE, two leading reverse-engineering projects. Both projects are based on the hypothesis of the columnar organization of the cortex. The rationale underlying reverse engineering in cognitive and brain sciences is outlined. New findings are mentioned questioning  the concept of the basic uniformity of the cortex, and consequences for the reverse-engineering projects are discussed. Section \ref{sec:7} ponders about the claim that non-decomposability is not an intrinsic property of complex systems but is only in our eyes, due to insufficient mathematical techniques. For this, the modeling relation as analyzed by Robert Rosen is explained which enables us to make the scientific analysis method itself a subject of discussion. It is concluded that the fundamental assumption of cognitive science must be abandoned. I end the paper by some conclusions  for the study of organisms and behavior as well as for engineering  artificial cognitive systems.

\section{Methodological Basics}
\label{sec:2}

\subsection{Decomposability}
\label{subsec:2.1}
Brains, even those of simple animals, are enormously complex structures, and it is a very ambitious goal to cope with this complexity. The scientific disciplines involved in cognitive and brain research are committed to a common methodology to explain the properties and capacities of complex systems. It is   decompositional analysis, i.e. analysis of the system in terms of its components or subsystems. 

\begin{figure}[ht]
\centering 
\includegraphics[width=0.7\textwidth]{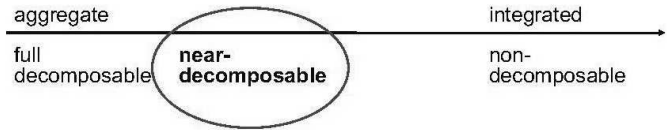}
\caption{Decomposability as a continously varying system property. According to this view, the focus is on near-decomposable systems which would represent the most relevant systems category in the cognitive and brain sciences.}
\label{fig:1} 
\end{figure}

Since Simon's influential book ``The Sciences of the Artificial'' \cite{Simon1969}, \mbox{(near-)} decomposability of complex systems  has been accepted as fundamental for the cognitive and  brain sciences. Cognitive capacities  are considered as dispositional properties which can be explained  via decompositional analysis. I call this the {\it fundamental assumption} for the cognitive and  brain sciences. Simon \cite{Simon1969}, Wimsatt \cite{Wimsatt1986} and Bechtel and Richardson \cite{BechtelRichardson1993}, among others, have further elaborated this concept. They consider decomposability  a continously varying system property, and state, roughly, that systems fall on a continuum from aggregate (full decomposable) to integrated (non-decomposable) (Fig. \ref{fig:1}). The {\it fundamental assumption} implies that cognitive and brain systems are  non-ideal aggregate systems; the capacities of the components are internally realized by strong intra-component interactions, and interactions between components do not appreciably contribute to the capacities; they are  much weaker than the intra-component interactions. Hence, the description of the complex system as a set of weakly interacting components seems to be a good approximation. This property of complex systems, which should have evolved through natural selection, was called near-decomposability \cite{Simon1969}.

Simon characterizes near-decomposability as follows: (1) In a nearly decomposable system, the short-run behaviour of each of the component subsystems is approximately independent of the short-run behaviour of the other components; (2) in the long run the behaviour of any one of the components depends in only an aggregate  way on the behaviour of the other components \cite[p.100]{Simon1969}. Thus, if the capacities of a near-decomposable system are to be explained, to some approximation its components can be studied in isolation, and based on their known interactions, their capacities eventually combined to generate the system´s behavior. 


Let us summarize this assumption because it is of central importance in the following:

\begin{ann}  \ \\ 
\vspace{-0.4cm}
\begin{quote}
Cognitive and brain systems are  non-ideal aggregate systems. The capacities of the components are internally realized (strong intra-component interactions) while interactions between components are negligible with respect to capacities. Any capacity of the whole system then results from superposition of the capacities of its subsystems. This property of cognitive and brain systems should have evolved through natural selection and is called near-decomposability.
\end{quote}
\end{ann}

\subsection{Decompositional Analysis}
\label{subsec:2.2}

The primary goal of cognitive science and its subdisciplines is to understand cognitive capacities like vision,  language, memory, planning etc. That is, we want to answer questions of the form "does system $S$ possess or exercise a capacity $C$?".  The quest for $S$'s capacity $C$ can be replaced by evaluating the proposition $P(S)$ = "$S$ possesses or exercises the capacity $C$". In other words, we want to determine the truth or falsity of the proposition $P(S)$.

Cummins \cite{Cummins1983,Cum2000} suggests that a system's capacity can be explained by a {\it functional analysis} of that capacity. A functional analysis of some capacity $C$ proceeds, roughly, by  parsing  the capacity into a set of constituent sub--capacities $C_{1}, C_{2}, . . ., C_{n}$.  Note that the sequence has to be specified in which those constituent capacities must be exercised for producing the complex capacity. That is, there is an algorithm which can be programmed to decide whether system $S$ has $C$ or $P$, by processing a finite list of propositions $P_{1}, P_{2},..., P_{n}$.

The scheme then  asserts that any capacity proposition $P(S)$ can be expressed as conjunction  of a finite number of sub-propositions $P_{i}(S)$ the truth of each one is necessary, and all together sufficient, for $P(S)$ to be true\footnote{Cummin's scheme evidently employs Frege's {\it principle of compositionality}, well-known in computer science as `divide and conquer'.}. Hence, a functional analysis comprises the following steps: 
\\

\noindent {\bf Functional analysis.}
\begin{enumerate}
	\item Establish that system $S$ has capacity $C$ or property $P$. 
	\item Decompose $P$ into sub-properties $P_{1}(S)$, $P_{2}(S)$,..., $P_{n}(S)$. 
	\item Specify the sequence in which the sub-properties $P_{i}$ are to be processed to 
generate $P$, i.e. the algorithm. 
\end{enumerate}


\begin{flushright} 
\rule{0.95\textwidth}{0.4pt}
\end{flushright}

\vspace*{-0.3cm}

\begin{eqnarray}
&&\mbox{Then it follows that $P(S)=\bigwedge^{n}_{i=1}P_{i}(S)$}.
\label{eqn:1}
\end{eqnarray}	

\vspace*{1cm}
If this scheme is applied to a material system $S$ with the property $P(S)$, it allows to express $P(S)$ in the form of eqn. (\ref{eqn:1}), i.e. by purely syntactical means. That is, property $P(S)$ is redundant, and its truth does not provide new information about system $S$, cf. \cite{Rosen2000}.

A cognitive capacity may be explained not only by analyzing the capacity itself, but also by analyzing the system that has it. This type  of decompositional analysis  is  {\it structural analysis} \cite{Atkinson1998,Eckardt2004}. It involves to attempt to identify the structural, material components of the system. Thus, the material system $S$ is to be decomposed into {\it context-independent} components $S_{j}$, i.e. their individual properties $P_{k}(S_{j})$ are independent of the decomposition process itself and of $S$'s environment.

\begin{figure}[ht]
\centering 
\includegraphics[width=0.6\textwidth]{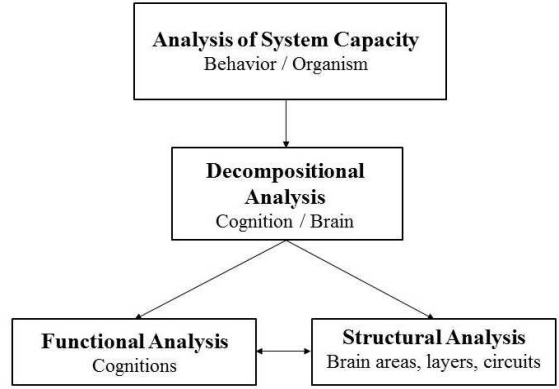}
\caption{View on decompositional analysis of brain and cognition. See text for details. }
\label{fig:2} 
\end{figure}

\subsection{Localization}
\label{subsec:2.4}
Functional analysis and structural analysis  must be clearly differentiated, although in practice, there is a close interplay between them (as indicated by the double arrow in Fig. \ref{fig:2}). This is obvious in the {\it localization approach} which combines both analysis types. The essential assumption is that each of the sub-properties $P_{i}(S)$, into which the property $P(S)$ was decomposed, is to be localized in some particular subsystem $S_{j}$ of $S$ itself. Thus, the properties $P_{k}(S_{j})$ of  $S$'s material components $S_{j}$ equal exactly the conjunction terms in eqn. (\ref{eqn:1}), i.e. for each sub-property $P_{i}(S)$ there is a structural component $S_{j}$ whose property $P_{k}(S_{j})$ is identical to $P_{i}(S)$, 

\begin{equation}
P_{i}(S)=P_{k}(S_{j}).
\label{eqn:2}
\end{equation}  

Thus, eqn.(\ref{eqn:1}) can be rewritten as

\begin{equation}
P(S)=\bigwedge_{j, k}P_{k}(S_{j}). 
\label{eqn:3}
\end{equation}

Equation (\ref{eqn:2})  in a nutshell expresses the idea of the {\it fundamental assumption}, i.e. decomposition and localization. Properties $P_{i}(S)$ of the whole system $S$ are identified with properties $P_{k}(S_{j})$ of certain of its subsystems $S_{j}$. This means, one assumes that any material system $S$ (including brain) can be decomposed into context-independent parts or structural components $S_{j}$ in such a way that their properties $P_{k}(S_{j})$ are independent of the properties of the other parts and of any environment. Thus, a set of decomposition operators ${\cal D}_{i}$ on $S$ of the form 

\begin{equation}
{\cal D}_{i}(S)=S_{i}
\label{eqn:4}
\end{equation} 

is supposed which isolate the subsystems $S_{i}$ from $S$. Corresponding to the {\it fundamental assumption}, the operators ${\cal D}_{i}$ break the inter--component interactions which `glue' the context-independent components of $S$ together, but without affecting any of the intra-component interactions. 

Ideally, decomposition operations like ${\cal D}_{i}$ are reversible, i.e. the whole system $S$ can be synthesized from the components $S_{j}$, 

\begin{equation}
S= S_{1}\otimes S_{2}\otimes ...\otimes S_{m}, 
\label{eqn:5}
\end{equation} 
   
where the $\otimes$-symbol denotes inter--component interactions like those broken by the decomposition operators ${\cal D}_{i}$. Thus, $S$ is to be considered as a kind of direct product. Now the close analogy of expressions (\ref{eqn:3})  and  (\ref{eqn:5}) becomes obvious: the fractionation of system $S$ corresponds to the compositionality of property $P(S)$ while the connector symbol $\wedge$  replaces  the inter--component interaction symbol $\otimes$,

\begin{equation}
P(S)=P_{1}(S_{1}) \wedge P_{2}(S_{2}) \wedge ...  \wedge P_{m}(S_{m}).
\label{eqn:6}
\end{equation}

These suppositions allow to proceed wholly in the syntactical realm. Any property $P$ of a physical system $S$ comes with an algorithm for evaluating $P$'s truth, and any physical system $S$ can be algorithmically generated from a sufficiently large population of components $S_{i}$ by exclusively syntactical means. In both cases, analysis and synthesis are inverse operations which are realized entirely by algorithms, i.e. the operations are computable, cf. \cite[p. 131]{Rosen2000}. 

Understating the case, the localization approach has been described as hypothetical identification which is to serve as research heuristics \cite{BechtelRichardson1993}. In fact, however, the majority of cognitive scientists considers it as fundamental and indispensable (e.g. \cite{Eckardt2004,Ross2010}). For example, Von Eckardt \cite{Eckardt2004} points out that a functional analysis for a capacity $C$ only provides us with a {\it possible} explanation of how the system has capacity $C$. That is because the decomposition of a composed capacity is not unique - it can be parsed into various alternative sequences of constituent capacities, each of which is sufficient for $S$'s capacity $C$. As a way out, she suggests to build a model that is structurally adequate by employing the localization approach. 

 A caveat is necessary, however: There is no reason to assume that functional and structural components match up one-to-one! Of course, it might be the case that some functional components map properly onto individual structural components. It is rather probable, however, for a certain functional component to be implemented by non-localized, spatially distributed material components.  Conversely, a given structural component may implement more than one distinct function. According to Dennett \cite[p. 273]{Dennett1991}: ``In a system as complex as the brain, there is likely to be much `multiple, superimposed functionality'.'' With other words, we cannot expect specific functions to be mapped to structurally bounded neuronal structures, and vice versa. It is now well known that Dennett's caveat has been proved as justified (e.g. \cite{PriceFriston2005}). Thus the value of the localization  approach as `research heuristics'  seems rather dubious \cite{Uttal2001,Henson2005}.

\subsection{Linearity, Modularization and Complex Systems}
\label{subsec:2.3}
 
In the cognitive and brain sciences, as in other fields of science, the components of  near-decomposable systems are called modules.  This term originates from engineering; it denotes the process of decomposing a product into building blocks - modules - with specified interfaces, driven by the designer's interests and intended functions of the product. It refers either to functional or structural components.  Modularized systems are linear in the sense that they obey an analog of the superposition principle  of linear system theory in engineering \cite{Schierwa1989}. If the modules are structurally localized functional components, the superposition principle is expressed by eqn. (\ref{eqn:5}). The function of a decomposable system results from the linear combination of the functions of the system modules\footnote{A corresponding class of models in mathematics is characterized by the superposition theorem for homogeneous linear differential equations stating that the sum of any two solutions is itself a solution.}. This principle mirrors the constructive step in the scheme of reverse engineering (see above and Section \ref{sec:5} below). The terms `linear' and `nonlinear' are often used in this way: `Linear' systems are decomposable into independent modules with linear, proportional interactions while `nonlinear' systems are not\footnote{We must differentiate between the natural, complex system and its description using modeling techniques from linear system theory or nonlinear mathematics.} \cite{Schierwa1989,Forrest1990}.  

Applying this concept to the systems at the other end of the complexity scale (Fig. \ref{fig:1}), the integrated systems  are basically not decomposable, due to the strong, nonlinear interactions involved. Thus, past or present states or actions of any or most subsystems always affect the state or action of any or most other subsystems. In practice, analyses of integrated systems nevertheless try to apply the methodology for decomposable systems, in particular if there is some hope that the interactions can be linearized. Such linearizable systems have been above denoted as nearly decomposable. However, in the case of strong nonlinear interactions, we must accept that decompositional analysis  is not applicable. 

Already several decades ago this insight was stressed. For example, Levins \cite[p.76 ff.]{Levins1970} around 1970 proposed a classification of systems into aggregate, composed and evolved systems. While the aggregate and the composed would not cause serious problems for decompositional analyses, Levins emphasized the special character of evolved systems:
\begin{quotation} {\it A third kind of system no longer permits this kind of analysis. This is a system in which the component subsystems have evolved together, and are not even obviously separable; in which it may be conceptually difficult to decide what are the really relevant component subsystems.... The decomposition of a complex system into subsystems can be done in many ways... it is no longer obvious what the proper subsystems are, but these may be processes, or physical subsets, or entities of a different kind.}  
\end{quotation} 

This statement clearly contradicts the {\it fundamental assumption}, and it has not lost its relevance, as the findings of complexity science have shown. Nevertheless, most researchers in the cognitive and brain sciences found reasons to cling to it. A main argument for the {\it fundamental assumption} has been that non--decomposability  is only in our eyes, and not an intrinsic property of strongly nonlinear systems, and scientific progress will provide us with the new mathematical techniques required to deal with nonlinear, integrated systems. I will return to this problem in Section \ref{sec:7}.

\section{Reverse Engineering the Brain}
\label{sec:5}
\subsection{The Column Concept}
\label{subsec:col}
\begin{figure}[ht]
\centering 
\includegraphics[width=0.5\textwidth]{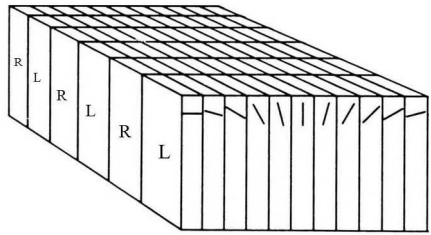}
\caption{Hubel and Wiesel's `ice cube' model of visual cortical processing. The diagram illustrates the idea that the cortex is composed of  iterated modules each of which  comprises a complete set of superimposed feature-processing elements, in this case for ocular dominance (indicated by L and R) and orientation selectivity (here represented for angles from 0 to $\pi$ ) (after \cite{HW1977}).}
\label{fig:4} 
\end{figure}

A guiding idea about the composition of the brain is the hypothesis of the columnar organization of the cerebral cortex. This {\it column concept} was developed mainly by Hubel and Wiesel, Mountcastle  and Szenth\'{a}gothai (e.g. \cite{HW1963,Mountc1997,Szenth1983}), and later on, it was published in the influential paper ``The basic uniformity in structure of the neocortex''  \cite{RHP}. According to this hypothesis (which has been taken more or less as fact by many experimental as well as theoretical neuroscientists), the neocortex is composed of `building blocks' (Fig. \ref {fig:4}) of repetitive structures, the `columns'  or `canonical cortical circuits', and it is characterized by a basic canonical pattern of  connectivity. In this scheme all cortical areas would perform identical or similar computational operations with their inputs. 

\subsection{Method of Reverse Engineering}
\label{subsec:rev}
Referring to and based on these works, several projects started recently, among them the {\it Blue Brain Project} \cite{Mark2006} and the {\it SyNAPSE Project} \cite{SyNAPSE}.  They are considered to be ``attempts to reverse-engineer the mammalian brain, in order to understand brain function and dysfunction through detailed simulations'' \cite{Mark2006} or, more pompous, ``to engineer the mind''\cite{SyNAPSE}. 

{\it Reverse engineering} is the main method used in empirical research to integrate the data derived from the different levels of the brain organization. Originally a concept in engineering and computer science, reverse engineering involves as first step a {\it decompositional analysis}, i.e. the detailed examination of a functional system ({\it functional analysis}) and its dissecting at the physical level into component parts ({\it structural analysis}), see Fig. (\ref{fig:2}). In a second step, the (re-) construction of the original system is attempted by creating duplicates including computer models, see below (Section \ref{sec:7}). This method is usually not much discussed with respect to its assumptions, conditions and range\footnote{Only recently, differences between proponents of reverse engineering on how it is appropriately to be accomplished became public. The heads of the two reverse engineering projects mentioned, Markram \cite{Mark2006} and Modha   \cite{SyNAPSE}, disputed publicly as to what granularity of the modeling is needed to reach a valid simulation of the brain. Markram questioned the authenticity of Modha's respective claims \cite{Brodkin2009}.} but see \cite{Dennett1994,Marom2009,Gurney2009}. 

The central role in these projects play cortical microcircuits or columns. As Maas and Markram \cite{Maass2004} formulate, it is a  ``tempting hypothesis regarding the computational role of cortical microcircuits ... that there exist genetically programmed stereotypical microcircuits that compute certain basis function.''  Their paper well illustrates the modular approach fostered, e.g. by \cite{Grillner2005,Gurney2009,Arbib1997,BresslerTognoli2006}. Invoking the localization concept, the tenet is that there exist fundamental correspondences among the anatomical structure of neuronal networks, their functions, and the dynamic patterning of their active states. Starting point is the `uniform cortex' with the cortical microcircuit or column  as the structural component. The question for the functional component is answered by assuming that there is a one-to-one relationship between the structural and the functional component (see Section \ref{subsec:2.2}). Together, the modularity hypothesis of the brain is considered to be both structurally and functionally well justified. 

As quoted above, the goal is to examine the hypothesis that there exist genetically programmed stereotypical microcircuits that compute certain basis function, thus providing for complex cognitive capacities. This hypothesis is based on the general, computational approach to cognitive capacities which takes for granted that ``cognition is computation'', i.e. the brain produces the cognitive capacities by computing functions\footnote{See \cite{Sch2007} for discussion of the computational approaches (including the neurocomputational one) to brain function, and their shortcomings.}. 

This assumption allows to apply the idea of decomposition or reverse engineering in the following way. From mathematical analysis and approximation theory it is well-known that a broad class of practically relevant functions $f$ can be approximated by composition or superposition of some basis functions. Of prime relevance in this respect are Kolmogorov's `superposition theorem' stating that continuous functions of $n$ arguments can always be represented using a finite composition of functions of a single argument, and addition, and Weierstrass and Stone's classical result that any real continuous function can be approximated with arbitrary precision using a finite number of computing units. Kolmogorov's theorem was rediscovered in the 1980s by Hecht-Nielsen and applied to artificial neural networks. Since then many different types of {\it function networks} and their properties have been investigated, so that a lot of results about the approximation properties of networks are already available, see e.g. \cite{Kolmo}.

\begin{figure}[ht]
\centering 
\includegraphics[width=0.8\textwidth]{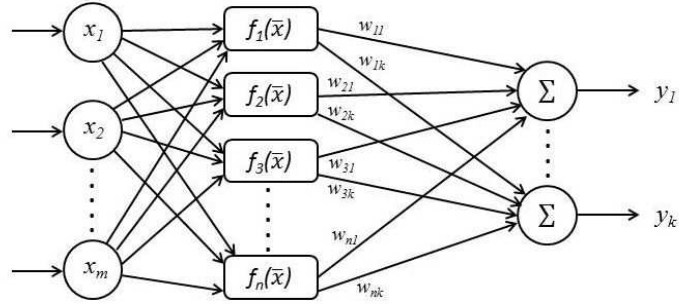}
\caption{Example of a network computing the function $f(\bar{x})=\bar{y}=(y_{1},...,y_{k})$.}
\label{fig:5} 
\end{figure}

For example, neural networks for function approximation have been developed based on orthogonal basis functions such as Fourier series, Bessel functions and Legendre polynomials. A typical configuration of such neural network of feedforward type  is illustrated in Fig. \ref{fig:5}. The input is $\bar{x}=(x_{1},...,(x_{m})$ and the output is $f(\bar{x})=\bar{y}=(y_{1},...,y_{k})$ with 

\begin{equation}
y_{j}=w_{1,j}\cdot f_{1}(\bar{x})+w_{2,j}\cdot f_{2}(\bar{x})+...+w_{n,j}\cdot f_{n}(\bar{x})
\label{eqn:7}
\end{equation}

The functions $f_{i}(\bar{x})  (i=1,...,n)$ are the basis functions computed by the network units. The real numbers $w_{i,j} \;(i=1,...,n; j=1,...,k)$ are their respective weights which can be adapted using effective learning algorithms to approximate the function $f(\bar{x})$.

As one can see, eqn. \ref{eqn:7} represents the analog of eqns. \ref{eqn:3} and \ref{eqn:5}, now in the computational realm. That is, from functional analysis and decomposition of a cognitive capacity into subcapacities, and fractionation of the cortex (or some subsystem) into parts we arrive at linear decomposition of a cognitive function into elements of a given set of `simple', or basis functions.

Thus, if some basis functions were identified, they provided the components of a (possible) computational decomposition. The reverse engineering method as applied in the cognitive and brain sciences from a computational perspective then proceeds as follows:

\begin{decoan} \ \\ 
\begin{enumerate}
	\item Capacity analysis: Specify a certain cognitive capacity which is assumed to be produced through the  cortex by computing a certain function. 
	\item Decompositional analysis: 
	\begin{enumerate}
	\item Functional (computational) analysis: Select a set of basis functions which might serve as functional components or computational units in the cortex.
	\item Structural analysis: Identify a set of anatomical components of the cortex. Provide evidence that cortical microcircuits are the anatomical components of the cortex.
	\end{enumerate}
	\item Localization: Provide evidence for the functional components or computational units being linked with the  anatomical components.\\  
	\item Synthesis: 
	\begin{enumerate}
	\item Modeling: 
		\begin{enumerate}
		\item Establish a structurally adequate functional model of the computational unit (the  presumed 'canonical circuit') which generates the  basis functions specified in step 2.(a). 
		\item Build a structurally adequate network model of the cortex (or some subsystem) composed of the canonical circuit models.
		\end{enumerate}
	\item Simulation: Prove that the specific cognitive capacity or function under study is generated by the network of circuit models, i.e through superposition of the specified basis functions.
	\end{enumerate}
\end{enumerate}
\end{decoan}

\subsection{Hypotheses and Reality} 
\label{subsec:7}

With the reverse engineering scheme formulated above, we have a 'recipe' at hand which could facilitate  the analysis very much. Recent findings in the experimental and theoretical fields, however, have objected most of the assumptions and hypotheses made as problematic, if not inappropriate and unrealistic. Already step 1, specification of a cognitive capacity, poses serious problems. It has always been extremely difficult to define exactly what is meant by a psychological, cognitive, or mental term, and the possibility should be acknowledged that they are only figments of our experimental designs or convenient artifices to organize our theoretical models \cite{Uttal2001}. This difficulty is obvious in recent attempts to build {\it cognitive ontologies} (e.g. \cite{PriceFriston2005,Henson2005}. 

Likewise, the assumptions about the structural and functional composition of the cortex, the notion of the basic uniformity in the cortex  with respect to the density and types of neurons per column for all species turned out to be untenable (e.g. \cite{HortonAdams2005,Rakic2009,Herculano2009,Fregnac2006}). It has been impossible to find the cortical microcircuit that computes a specific basis function \cite{deGaris2010,Fregnac2006}. No genetic mechanism has been deciphered that designates how to construct a column. The column structures encountered in many species (but not in all) seem to represent spandrels (structures that arise non-adaptively, i.e. as an epiphenomenon) in various stages of evolution \cite{Gould}.

Step 4  - synthesis - is worth extended discussion which space limitations forbid. In short, this step represents the conviction that large-scale modeling of brain networks will eventually lead to understanding the mind-brain problem. It has been argued that producing and understanding complex phenomena from the interaction of simple nonlinear elements like artificial neurons or cellular automata  is possible. One expects then, that this would also work for cortical circuits which are recognized as nonlinear devices, and theories could be applied (or developed, if not yet available) that would guide us to which model setup might have generated a given network behavior. This would complete the reverse engineering process.

However, findings in complexity or nonlinear science exclude such transfer of the usual, linear approach. It is now clear that finding out which processes caused a specific complex behavior of a given system - an {\it inverse problem} - is hard because of its {\it ill-posedness}\footnote{In mathematics, a problem is called ill-posed if no solution or more than one solution exists, or if the solutions depend discontinuously upon the initial data.}. This means for the study of cortical circuits and networks of them that from observed activity or function we cannot, in principle, infer the internal organization. A wide variety of different organizations can produce the same behavior.

If we revisit the column concept of the cortex employed in theories of brain organization, we recognize that hypothesized structural components (cortical columns) have been identified with alike hypothetical functional components (basis function), employing the localization concept (Section \ref{subsec:2.2}). As we have seen, the facts contradict these assumptions, i.e. the reverse engineering project has been based on false presuppositions. In contrast to the localization idea, there is evidence for a given functional component to be implemented by spatially distributed networks and, vice versa, for a given structural component to implement more than one distinct function. With other words, it is not feasible for  specific functions to be mapped to structurally bounded neuronal structures \cite{PriceFriston2005,HortonAdams2005,Rakic2009,Herculano2009}.

This means, although the column concept is an attractive idea both from neurobiological and computational point of view, it cannot be used as an unifying principle for understanding cortical function. Thus, it has been concluded  that the concept of the cortex as a `large network of identical units' should be replaced with the idea that the cortex consists of `large networks of diverse elements' whose cellular and synaptic diversity is important for computation (e.g. \cite{Fregnac2006}.

It is worth to notice that the claim for conceptual change towards `cortex as large network of diverse elements' completely remains within the framework of reverse engineering, i.e. it is a plea for `Just carry on!'. It appears questionable, however, that the original goals of the cognitive and brain sciences and AI can be achieved this way. Actually, the methods of decompositional analysis and reverse engineering themselves have been principally criticized, which will be shortly discussed in the next section.

\section{The Modeling Relation}
\label{sec:7}

In Section \ref{subsec:2.3}, I concluded that complex, integrated systems are basically non-decomposable, thus resisting the standard analysis method. Now I return to this issue and to the consequences for investigating such systems in the cognitive and brain sciences. 

Despite contradicting findings in complex systems science, the majority of researchers in the cognitive and brain sciences subscribes for the {\it fundamental assumption}, i.e. the relevant systems in the cognitive and brain sciences are treated as nearly decomposable. Accordingly, non-decomposability is considered not as intrinsic property of complex, integrated systems but only as subjective, temporary failure of our methodology, due to insufficient mathematical techniques (e.g. \cite{Bechtel_2002}). 

\begin{figure}[ht]
\centering 
\includegraphics[width=0.75\textwidth]{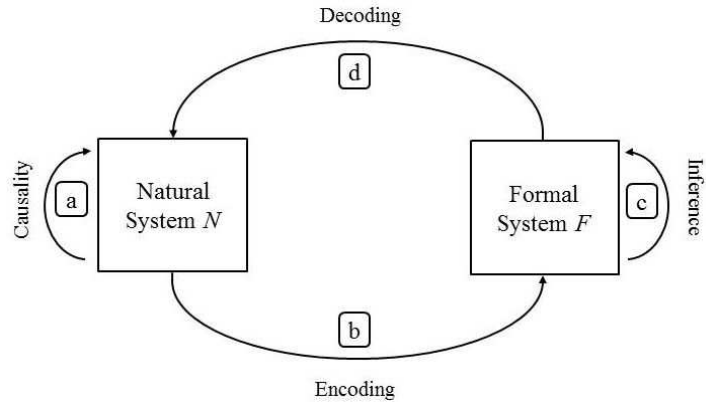}
\caption{The modeling relation. A natural system $N$ is modeled by a formal system $F$. Each system has its own internal entailment structures (arrows a and c), and the two systems are connected by the encoding and decoding processes (arrows b and d). After \cite[p.60]{Rosen1991}.}
\label{fig:3} 
\end{figure}

In contrast to that, Rosen \cite{Rosen1991,Rosen2000} has argued that understanding complex, integrated systems requires making  the scientific analysis method itself  a subject of discussion.  A powerful method of understanding and exploring the nature of the scientific method, and in particular, reverse engineering, provides the {\it modeling relation}. It is this relation by which scientists bring  ``entailment structures into congruence'' \cite[p. 152]{Rosen1991}. The modeling relation is represented by the set of mappings shown in Fig. \ref{fig:3}. It relates two systems, a natural system $N$ and a formal system $F$, by a set of arrows depicting processes or mappings. The assumption is that this diagram represents the various processes which we are carrying out when we perceive the world.

The  modeling relation is a relation in the formal mathematical sense, 

\begin{equation}
R=\left\{(a, c) \;| \; a=b \circ c \circ d\right\}
\end{equation}

while $\circ$ denotes concatenation. The members $a$ and $c$ of each ordered pair in $R$ are entailments from the two systems, $N$ and  $F$. Natural system $N$ is  part of the physical world that we wish to understand (in our case: human being, organism, brain), in which things happen according to rules of causality (arrow $a$). That is,  if some cause  acts on $N$, then the system will behave in a certain way, or produce certain effects. This resultant coupling of cause and effect in  $N$  is called  {\it causal entailment}. 

On the right of Fig. \ref{fig:3}, $F$ represents symbolically the parts of  the natural system (observables) which we are interested in, along with formal rules of inference (arrow $c$) that essentially constitute our working hypotheses about the way things work in $N$, i.e. the way in which we manipulate the formal system to try to mimic causal events observed or hypothesized in the natural system on the left. Stated another way, $F$  has inferential linkage; that is, if some premise proposition acts on $F$, then it  will generate  a consequential proposition as conclusion.  This resultant coupling of premise and conclusion in $F$ is called {\it  inferential entailment}.

Arrow $b$ represents the encoding of the parts of $N$ under study into the formal system $F$, i.e. a mapping that establishes the correspondence between observables of $N$ and symbols defined in $F$. Predictions about the behavior in $F$, according to $F$'s rules of inference, are compared to observables in $N$ through a decoding represented by arrow $d$. When the predictions match the observations on $N$, we say that $F$ is a successful model for $N$. Otherwise the entailment structures could not be brought into congruence, thus $F$ failed to model $N$.

It is important to note that the encoding and decoding mappings are independent of the formal and natural systems, respectively. In other words, there is no way to arrive at them from within the formal system or natural system. That is, the act of modeling is really the act of relating two systems in a subjective way. That relation is at the level of observables; specifically, observables which are selected by the modeler as worthy of study or interest. 

Given the modeling relation and the detailed structural correspondence between our percepts and the formal systems into which we encode them, it is possible to make a dichotomous classification of systems into those that are {\it simple} or {\it predicative} and those that are {\it complex} or {\it impredicative}. This classification can refer to formal inferential systems such as mathematics or logic, as well as to physical systems. As Rosen showed \cite{Rosen1991,Rosen2000}, a simple system is one that is definable completely by algorithmic method: All the models of such a system are Turing-computable or simulable. When a single dynamical description is capable of successfully modeling a system, then the behaviors of that system will, by definition, always be correctly predicted. Hence, such a system will be {\it predicative} in the sense that there will exist no unexpected or unanticipated behavior.

A complex system is by exclusion not a member of the syntactic, algorithmic class of systems. Its main characteristics are as follows. A complex system possesses non-computable models; it has inherent impredicative loops in it. This means, it requires multiple partial dynamical descriptions - no one of which, or combination of which, suffices to successfully describe the system. It is not a purely syntactic system as described by eqns. (\ref{eqn:1}) -- (\ref{eqn:5}) but it necessarily includes semantic elements. Complex systems also differ from simple ones in that complex systems cannot be linearly composed of parts - they are non-decomposable. This means, when a complex system is decomposed, its essential nature is broken by breaking its impredicative loops.

This has important effects. Decompositional analysis is inherently destructive to what makes the system complex - such a system is not decomposable without losing the essential nature of the complexity of the original system! In addition, by being not decomposable, complex systems no longer have analysis and synthesis as simple inverses of each other. Building a complex system is therefore not simply the inverse of any analytic process of decomposition into parts, i.e. the system is not a direct product of components, thus eqn. (\ref{eqn:5}) does not  hold.  

Since the brain is a complex, integrated and thus non-decomposable system, both steps of reverse engineering  -- decomposition into functional and structural components and subsequent synthesis -- must necessarily fail and will not provide the envisaged understanding!

It should be stressed that simple and complex systems after Rosen's definition cannot be directly related to those sensu Simon (Section \ref{sec:2}). While Rosen's approach yields a {\it descriptive} definition of complexity, Simon's is {\it interactional}, see \cite{Wimsatt1972}. It seems clear, however, that Rosen's `simple systems' comprise Simon's full- and near-decomposable systems, and Rosen's `complex systems' correspond to Simon's non-decomposable, integrated systems, as well as to Levin's evolved systems. No matter which definition is applied, the conclusion about the brain's non--decomposability remains valid.

\section{Conclusions}
\label{sec:8}
If one attempts to understand a complex system like the brain it is of crucial importance if general operation principles can be formulated. Traditionally, approaches to reveal such principles follow the line of  decompositional analysis as expressed in the {\it fundamental assumption} of cognitive and  brain sciences, i.e.  cognitive systems like other, truly complex systems are decomposable. Correspondingly, reverse engineering has been considered the appropriate methodology to understand the brain and to engineer artificial cognitive systems.   

I have argued that this position is at odds with the findings of complexity science. In fact, non-decomposability {\bf {\it is}} an intrinsic property of complex, integrated systems, and cannot be considered as subjective, temporary failure of our methodology, due to insufficient mathematical techniques. Thus, the dominant complexity concept  of cognitive and brain sciences underlying reverse engineering needs revision. The updated, revised concept must comprise results from the nonlinear science of complexity  and insights expressed, e.g., in Rosen's work on life and cognition. In the first line, this means that the {\it fundamental assumption} of cognitive and brain sciences must be abandoned. 

Organisms and brains are complex, integrated systems which are non--decomposable. This insight implies that there is no `natural'  way to decompose the brain, neither structurally nor functionally. We must face the uncomfortable insight that in cognitive and brain sciences we have conceptually, theoretically, and empirically to deal  with complex, integrated systems which is much more difficult than with simple, decomposable systems of quasi--independent modules! Thus, we cannot avoid (at least in the long run) subjecting research goals such as the creation of `brain-like intelligence' and the like to analyses which apprehend the very nature of natural complex systems.


\begin{thebibliography}{00}

\bibitem{Sch2007} Schierwagen, A.: Brain Organization and Computation. In: J. Mira and J.R. Alvarez (eds.) IWINAC 2007, Part I: Bio-inspired Modeling of Cognitive Tasks, LNCS 4527, pp. 31-–40 (2007)  
\bibitem{Sch2009}Schierwagen, A.: Brain Complexity: Analysis, Models and Limits of Understanding.  In: J. Mira et al. (Eds.): IWINAC 2009, Part I, LNCS 5601, pp. 195–-204 (2009)

\bibitem{Brooks2001} Brooks, R.: The relationship between matter and life. Nature 409 (2001)  409–-410 
\bibitem{Novere2007} Le Novere, N.: The long journey to a Systems Biology of neuronal function. BMC Syst Biol. (2007)1--28.
\bibitem{Grillner2005} Grillner, S., Markram,H.,  De Schutter,E., Silberberg, G., LeBeau,F.E.N.: Microcircuits in action – from CPGs to neocortex. Trends in Neurosciences 28 (2005) 525--533

\bibitem{Brodkin2009} Brodkin, J.:  IBM cat brain simulation dismissed as `hoax' by rival scientist. Network World, November 24 (2009)  

\bibitem{Dennett1994} Dennett, D.C.: Cognitive Science as Reverse Engineering: Several Meanings of 'Top Down' and 'Bottom Up,'. In D. Prawitz, B. Skyrms, D. Westerståhl (eds.)  Logic, Methodology and Philosophy of Science IX, pp. 679--689. Elsevier Science,  Amsterdam (1994) 
\bibitem{Marom2009} Marom, S., Meir, R., Braun, E., Gal, A., Kermany, E., Eytan, D.: On the precarious path of reverse neuro-engineering. Front. Comput. Neurosci. 3  (2009) doi:10.3389/neuro.10.005.   
\bibitem{Gurney2009} Gurney, K.: Reverse engineering the vertebrate brain: methodological principles for a biologically grounded programme of cognitive modelling.  Cognit. Computat. {\bf 1},  29--41 (2009)


\bibitem{Rosen1991} Rosen, R.: Life Itself: A Comprehensive Inquiry into the Nature, Origin, and Fabrication of Life. Columbia University Press, New York (1991) 
\bibitem{Rosen2000} Rosen, R.:  Essays on Life Itself. Columbia University Press, New York (2000)

\bibitem{Simon1969} Simon. H.: The Sciences of the Artificial. MIT Press, Cambridge, MA (1969)

\bibitem{Wimsatt1986} Wimsatt, W.: Forms of aggregativity. In: A. Donagan, A.N. Perovich, M.V. Wedin (eds.) Human Nature and Natural Knowledge, pp. 259—-291.  D. Reidel, Dordrecht (1986) 
\bibitem{BechtelRichardson1993} Bechtel, W., Richardson,  R. C.:  Discovering complexity: Decomposition and 
        localization as strategies in scientific research. Princeton University Press, Princeton, NJ (1993)


\bibitem{Cummins1983} Cummins, R.: The Nature of Psychological Explanation. MIT Press, Cambridge, MA (1983)
\bibitem{Cum2000} Cummins, R.: "How Does It Work" Versus "What Are the Laws?": Two Conceptions of Psychological Explanation. In: F. Keil, R. A. Wilson (eds.) Explanation and Cognition, pp. 117--145. MIT Press, Cambridge, MA (2000)
\bibitem{Atkinson1998} Atkinson, A.P.: Persons, systems and subsystems: The explanatory scope of cognitive psychology. Acta Analytica {\bf 20}, 43--60 (1998) 

\bibitem{Eckardt2004} Eckardt, B. von, Poland, J.S. :  Mechanism and Explanation in Cognitive Neuroscience. Philosophy of Science, {\bf71}, 972--984 (2004)


\bibitem{Dennett1991}Dennett, D. C.: Consciousness explained. Little, Brown and Co, Boston (1991)
\bibitem{PriceFriston2005} Price, C.J., Friston, K.J.: Functional ontologies for cognition: the systematic definition of structure and function. Cogn. Neuropsychol {\bf 22}, 262-–275 (2005)
\bibitem{Uttal2001} Uttal, W. R.: The New Phrenology. The Limits of Localizing Cognitive Processes in the Brain. Cambridge, Mass., MIT Press (2001)
\bibitem{Henson2005} Henson, R.:  What can functional neuroimaging tell the experimental psychologist?  Quart. J. Exper. Psychol. {\bf 58A},  193--233 (2005)
\bibitem{Ross2010} Ross, E.D.: Cerebral Localization of Functions and the Neurology of Language: Fact versus Fiction or Is It Something Else? Neuroscientist {\bf 16}, 222--243 (2010)


\bibitem{Schierwa1989} Schierwagen, A.: Real neurons and their circuitry: Implications for brain theory. iir--reporte, pp. 17--20. Akademie der Wissenschaften der DDR, Institut f\"{u}r Informatik und Rechentechnik), Eberswalde (1989) 
\bibitem{Forrest1990}  Forrest, S.: Emergent computation : self-organizing, collective, and cooperative phenomena in natural and artificial computing networks. Physica D {\bf 42}, 1--11 (1990)

\bibitem{Levins1970} Levins, R.: Complex Systems. In: C. H. Waddington (ed.) Towards a Theoretical Biology, Vol. 3, pp. 73--88. University of Edinburgh Press, Edinburgh (1970)  

\bibitem{Edmonds2009} Edmonds, B.: Understanding Observed Complex Systems – the hard complexity problem. CPM Report No.: 09-203 (2009)
\bibitem{HW1977} Hubel, D. H., Wiesel, T. N.:   Ferrier Lecture: Functional Architecture of Macaque Monkey Visual Cortex. Proc. R. Soc. Lond. B {\bf 198}, 1-59 (1977)

\bibitem{Mark2006} Markram, H.: The Blue Brain Project. Nature Rev. Neurosci. {\bf 7}, 153--160 (2006) 

\bibitem{SyNAPSE} Systems of Neuromorphic Adaptive Plastic Scalable Electronics (SyNAPSE).  DARPA / IBM (2008)        


\bibitem{HW1963} Hubel, D. H., Wiesel, T. N.:  Shape and arrangement of columns in cat's striate cortex. J. Physiol. {\bf 165}, 559–-568  (1963)
\bibitem{Mountc1997} Mountcastle, V. B.: The columnar organization of the neocortex. Brain {\bf 120},  701-–722 (1997)
\bibitem{Szenth1983} Szenth\'{a}gothai J.: The modular architectonic principle of neural centers. Rev. Physiol. Bioche.  Pharmacol. {\bf 98},  11--61 (1983)
\bibitem{RHP}Rockel A.J., Hiorns R.W., Powell T.P.S.: The basic uniformity in structure of the neocortex. Brain {\bf 103},  221–-244 (1980)


\bibitem{Maass2004} Maass, W., Markram, H.: Theory of the computational function of microcircuit dynamics. In: S. Grillner and A. M. Graybiel   (eds.) The Interface between Neurons and Global Brain Function, Dahlem Workshop Report 93, pp. 371--390.  MIT Press, Cambridge, MA (2006) 

\bibitem{Arbib1997} Arbib, M., \'{E}rdi, P., Szenth\'{a}gothai, J.: Neural Organization: Structure, Function and Dynamics. MIT Press, Cambridge, MA (1997)

\bibitem{BresslerTognoli2006}  Bressler,S.L., Tognoli,E.: Operational principles of neurocognitive networks, Intern. J. Psychophysiol. {\bf 60},  139–-148 (2006)

\bibitem{Kolmo} Suykens, J.A.K., Vandewalle, J.P.L., Moor, B.L. de:  Artificial Neural Networks for Modelling and Control of Non-Linear Systems. Kluwer Academic Publishers, Dordrecht (1996)


\bibitem{HortonAdams2005}  Horton, J. C., Adams, D. L.: The cortical column: a structure without a function. Phil. Trans. R. Soc. B {\bf 360}, 386--62  (2005) 
\bibitem{Rakic2009}Rakic, P.: Confusing cortical columns.  Proc. Natl. Acad. Sci. USA {\bf 105}, 12099--12100 (2008)
\bibitem{Herculano2009} Herculano-Housel, S., Collins, C.E., Wang, P., Kaas, J.: The basic nonuniformity of the cerebral cortex. Proc. Natl. Acad. Sci. USA {\bf 105}, 12593-–12598 (2008)
\bibitem{Fregnac2006} Fr\'{e}gnac, Y. et al.: Ups and downs in the genesis of cortical computation. In: S. Grillner and A.M. Graybiel (eds.) Microcircuits: The Interface between Neurons and Global Brain Function, Dahlem Workshop Report 93, pp. 397--437.   MIT Press, Cambridge, MA  (2006)  
\bibitem{deGaris2010} de Garis, H., Shuo, C., Goertzel, B. and Ruiting, L.: A world survey of artificial brain projects PartI: Large-scale brain simulations, Neurocomputing, doi:10.1016/j.neucom.2010.08.004 


\bibitem{Gould} Gould, S.J., Lewontin, R.C.: The spandrels of San Marco and the Panglossian paradigm: a critique of the adaptationist programme. Proc. Roy. Soc. London B {\bf 205}, 581--598 (1979)

\bibitem{Bechtel_2002} Bechtel, W.: Decomposing the brain: A long term pursuit. Brain and Mind {\bf 3}, 229--242 (2002)

\bibitem{Wimsatt1972} Wimsatt, W.C.: Complexity and Organization. Proc. Biennial Meeting Philos. Sci. Ass. {\bf 1972},  67--86 (1972)

\end{thebibliography}
\end{document}